\documentclass[%
 aip,
 amsmath,amssymb,
 reprint,%
]{revtex4-1}

\usepackage{graphicx}
\usepackage{dcolumn}
\usepackage{bm}

\usepackage[utf8]{inputenc}
\usepackage[T1]{fontenc}
\usepackage{mathptmx}
\usepackage{etoolbox}

\makeatletter
\def\@email#1#2{%
 \endgroup
 \patchcmd{\titleblock@produce}
  {\frontmatter@RRAPformat}
  {\frontmatter@RRAPformat{\produce@RRAP{*#1\href{mailto:#2}{#2}}}\frontmatter@RRAPformat}
  {}{}
}%
\makeatother

\usepackage{centernot}
\usepackage{graphicx}
\usepackage{amsmath}
\usepackage{amsbsy}
\usepackage{times}
\usepackage{amssymb}
\usepackage{mathrsfs}
\usepackage{chemarr}
\usepackage[dvipsnames]{xcolor}
\usepackage{url}
\usepackage{upgreek}
\usepackage{version}
\usepackage[normalem]{ulem}
\usepackage[pdftex,colorlinks=true,pdfstartview=FitV,linkcolor= linkcolor,citecolor= linkcolor,urlcolor= linkcolor,hyperindex=true,hyperfigures=false]{hyperref}
\definecolor{linkcolor}{rgb}{0,0,0.6}
\usepackage{lipsum}
\usepackage{printlen}
\usepackage{tikz}
\usepackage{accents}

\usepackage{siunitx}
\usepackage[version=4]{mhchem}

\newcommand\bj{{\bf j}}

\newcommand\br{{\bf r}}

\newcommand\bv{{\bf v}}
\newcommand\bu{{\bf u}}

\newcommand\bF{{\bf F}}

\definecolor{redUP}{RGB}{138,21,56}

\definecolor{gianmarco}{HTML}{ed2867}

\definecolor{hdacolor}{HTML}{1A6FA3}

\begin{document}

\preprint{APS/123-QED}

\title{%
Closing of density wake in concentrated colloidal suspensions:\linebreak
two regimes governed by compressibility, friction, and self-propulsion
}

\author{Guillaume Duprez}
\thanks{GD and GS contributed equally to this work.}
\affiliation{
 Université Lyon 1, CNRS, ILM, UMR 5306, Villeurbanne, France
}

\author{Gianmarco Spera}
\thanks{GD and GS contributed equally to this work.}
\affiliation{
Rudolf Peierls Centre for Theoretical Physics, University of Oxford, Oxford OX1 3PU, United Kingdom
}
\author{Benjamin Buhl}
\affiliation{
 Université Lyon 1, CNRS, ILM, UMR 5306, Villeurbanne, France
}

\author{François Graner}
\affiliation{Université Paris Cité, CNRS, Matière et systèmes complexes, F-75013 Paris, France}

\author{Vincent Démery}
\author{Cécile Cottin-Bizonne}
\thanks{CCB, HDA, and ML equally supervised this work.}
\email{cecile.cottin-bizonne@univ-lyon1.fr}
\author{Hélène Delanoë-Ayari}
\thanks{CCB, HDA, and ML equally supervised this work.}
\email{helene.delanoe-ayarih@univ-lyon1.fr}
\author{Mathieu Leocmach}
\thanks{CCB, HDA, and ML equally supervised this work.}
\email{mathieu.leocmach@univ-lyon1.fr}
\affiliation{
 Université Lyon 1, CNRS, ILM, UMR 5306, Villeurbanne, France
}

\date{\today}

\begin{abstract}
When an obstacle moves through a compressible medium, it creates a density perturbation. Theoretical arguments regarding the shape of this density wake in Brownian suspensions either suppose infinite dilution or describe only the far field. Here, we study via experiments and numerical simulations the closing of the density perturbation just behind a circular obstacle in dense 2D suspensions of particles that are thermal or self-propelled. In the thermal case, we show that the characteristic length of the wake is a non-monotonic function of density, decreasing then increasing. We rationalise the scalings of both regimes by analytical arguments that highlight the role of the compressibility factor of the suspension. Based on this new understanding of the passive case, we show that the influence of self-propulsion observed in experiments and simulations can be understood as if the obstacle is effectively larger in an active bath, with non-trivial and distinct roles of the self-propulsion velocity and of the persistence time.
\end{abstract}

\maketitle

\section{Introduction}
The flow around an obstacle is a classical experiment in fluid dynamics since Stokes~\cite{Stokes1850}. It is also a test bed for complex fluids~\cite{leal1988LakeArrowheadWorkshop, brown1994ReportVIIIthInternational, alves2021NumericalMethodsViscoelastic, hawardBifurcationsFlowsComplex}. 
Its spatially heterogeneous geometry enables to probe and discriminate complex rheologies~\cite{cheddadi2011UnderstandingPredictingViscous} with more efficiency than classical rheometric flows such as Couette~\cite{cheddadi2012NonUnique}.
In particular viscoelastic and viscoplastic fluids can display asymmetric flow field around a symmetric obstacle, unsteady flow~\cite{hopkins2022UpstreamWallVortices, hawardBifurcationsFlowsComplex}, and complex recirculations~\cite{arigo1998ExperimentalInvestigationNegative}. 
Viscoelastoplastic flows display features qualitatively different from viscoelastic and viscoplastic ones~\cite{cheddadi2011UnderstandingPredictingViscous}.
In granular media~\cite{chehata2003DenseGranularFlow, candelier2010JourneyIntruderFluidization, kolb2013RigidIntruderTwodimensional, seguin2016LocalRheologicalMeasurements} and suspensions~\cite{carpen2005MicrorheologyColloidalDispersions, squires2005simple, meyer2006LaserTweezerMicrorheology, sriram2010ActiveMicrorheologyColloidal}, 
moving an obstacle through the fluid also perturbs the particle density field, with a lower density region downstream and a higher density region upstream.
Understanding the shape of this perturbation is essential to work out the mechanical response of the compressible medium to the obstacle motion~\cite{squires2005simple}.
Furthermore, how the density wake opens and closes is of interest by itself, as it reveals the dynamical processes at play in the system and how emerging properties like pressure, viscosity or particle mobility couple with density.

The existence of a density wake stems from finite compressibility of the medium -- particle concentration can be heterogeneous -- and excluded-volume interactions between the obstacle and the particles.
However, the closing of the density perturbation can have various physical origins.
In the absence of thermal motion or external vibration, granular particles do not diffuse and interact only at contact. In the absence of gravity, as for instance in a 2D horizontal granular layer, the wake is an empty cavity behind the obstacle which area is limited only by the possible extent of the jammed zone in front of the obstacle~\cite{kolb2013RigidIntruderTwodimensional}.
A density wake can thus exist even in the case of quasi-static motion of the probe.
By contrast, in a granular pile the density wake is a dynamic phenomenon that balances obstacle velocity with rearrangement processes due to gravity~\cite{rognon2025DragForcesGranular}. Whether gravity effects are just the local free fall of grains on the obstacle size or a more global phenomenon akin to hydrostatic pressure is still debated~\cite{rognon2025DragForcesGranular} and might depend on the history of the pile~\cite{guillard2013DepthIndependentDragForce}.

In a dilute thermal suspension of hard spheres or hard disks, particle diffusion balances the density perturbation caused by obstacle motion~\cite{squires2005simple}. This balance is summed up by the Péclet number $\mathrm{Pe}_\mathrm{dilute} = U_\mathrm{obs}R/D_\mathrm{t}$, where $U_\mathrm{obs}$ is the velocity of the obstacle, $D_\mathrm{t}$ the translational diffusion coefficient of the particles, and $R=R_\mathrm{obs}+a$ is the radius of the zone from which particle centres are excluded around the obstacle, with $R_\mathrm{obs}$ and $a$ the respective radii of the obstacle and a particle.
At low Péclet the wake is a dipolar perturbation of the density field, decaying as $\propto r^{-1}$ in 2D, whereas at high Péclet the density becomes null behind the obstacle while a one particle-thick dense layer forms in front.
Extensions of this thermal picture have been proposed for dilute suspensions of self-propelled particles~\cite{burkholder2020NonlinearMicrorheologyActivea}. When the persistence length $\ell$ of the particle motion is similar to $R$, the effect of self-propulsion cannot be described as an effective diffusion, and a ballistic closing of the wake is observed~\cite{burkholder2020NonlinearMicrorheologyActivea, knezevic2021OscillatoryActiveMicrorheologya}.
However, the effect of density on wake closing remains unclear, both for passive and active suspensions.

Indeed, most studies on obstacles in suspension of thermal or self-propelled particles have a microrheology focus: what is the drag force given the velocity of the probe, or what is the velocity given the applied force~\cite{squires2005simple,burkholder2020NonlinearMicrorheologyActivea, knezevic2021OscillatoryActiveMicrorheologya, Puertas2014, reichhardt2015ActiveMicrorheologyActive, peng2022ForcedMicrorheologyActive}. Beyond the dilute regime, a simple replacement of $D_\mathrm{t}$ by the collective diffusion coefficient in the expression of $\mathrm{Pe}^\mathrm{dilute}$ describes well microrheological responses~\cite{sriram2010ActiveMicrorheologyColloidal}.
Density wakes have been observed in simulations~\cite{carpen2005MicrorheologyColloidalDispersions, weeber2012HydrodynamicInteractionsActive} and experiments~\cite{meyer2006LaserTweezerMicrorheology, sriram2010ActiveMicrorheologyColloidal} of dense thermal suspensions but their shapes have not been quantified.
Hydrodynamic models have predicted power law decay ($\propto r^{-3/2}$ in 2D) of the density perturbation at long range in dense assemblies of soft~\cite{demery2014generalized} and hard colloids~\cite{banerjee2022unjamming}, but do not describe the short range perturbation.

In this article, we address the question of wake closing in the context of dense passive and active colloidal systems by means of experiments, simulations, and continuum models. 
Experimentally, we consider an obstacle displaced at constant velocity in a monolayer of Janus colloidal particles that can be made self-propelled.
We show that the density wake behind the obstacle has a characteristic length at short range. 
In the passive case, we characterise the dependence of this length with density and its scaling with obstacle velocity and size.
Next, we characterize further the dependencies of this length in using numerical simulations of passive Brownian particles.
We discover two regimes in density which we explain by introducing a passive continuum model that brings out the physical ingredients underlying our experimental observations: compressibility and friction.
We then study how self-propulsion affects this scenario, both in simulations and in experiments, distinguishing between the respective effect of velocity and persistence time.
Finally, we discuss how our new understanding of the density wake behind an obstacle can help the characterisation of passive or active suspensions.
 
\begin{figure}
    \includegraphics{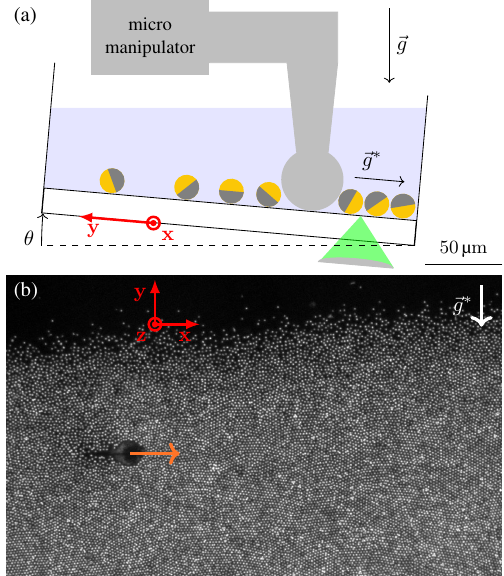}
    \caption{
        \textbf{(a)} Sketch of the experimental setup. Janus colloids form a monolayer that is sedimented by tilting the experimental setup by a small angle $\theta$. A micromanipulator holds the obstacle that dips in the monolayer and is dragged at constant velocity along $\vec{x}$. 
        \textbf{(b)} Detail of a dark-field microscopy image of the sedimented monolayer in the passive case. In-plane gravity $\vec{g}^*$ acts along $\vec{y}$. The orange arrow represents the velocity at which the obstacle is dragged (from left to right, in all figures). The obstacle in motion appears as a circular zone depleted in colloids. The density wake is visible as a low-density area behind the obstacle. 
    }
    \label{fig:xp}
\end{figure}

\begin{figure}
    \includegraphics[width=\columnwidth]{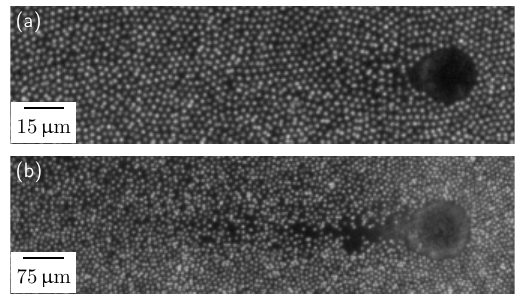}

    \caption{Experimental micrograph of the wake for two different obstacle radii: (a)  $R=\SI{8.5}{\micro\metre}$ and (b) $R=\SI{42}{\micro\metre}$, at roughly the same density ($\phi=0.767\pm 0.025$ and $0.758\pm 0.031$), and the same velocity $U_\mathrm{obs}=\SI{5}{\micro\metre/\second}$. The two pictures are scaled so that $R$ appears the same to highlight the nonlinear dependence of wake length on obstacle size.}
    \label{fig:xp_size_dependency}
\end{figure}

\section{Experimental setup}
As in Ref~\cite{finscarreira2024HowSteerActive}, we study $\SI{0.8}{\micro\metre}$-radius gold particles grafted with octadecanethiol~\cite{liSelfPropelledNanomotorsAutonomously2015} and half-coated with Platinum. After suspension in pure water, the particles immediately sediment onto the flat bottom of the experimental cell, forming a bidimensional monolayer.
The system is observed through a 20x fluotar objective of a Leica DMI 4000B microscope equipped with a custom-made external dark-field illumination and a camera (Basler ac-A2040-90um) recording at \SI{10}{fps}.
As shown in Fig.~\ref{fig:xp}(a), a very low in-plane apparent gravity $\vec{g}^*$  
is obtained by tilting the whole set-up with a small angle $\theta \approx \ang{0.1}$. We thus obtain a 2D sediment within the monolayer, with a density gradient along $y$, as shown in Fig.~\ref{fig:xp}(b).

\begin{figure*}
    \centering
    \includegraphics{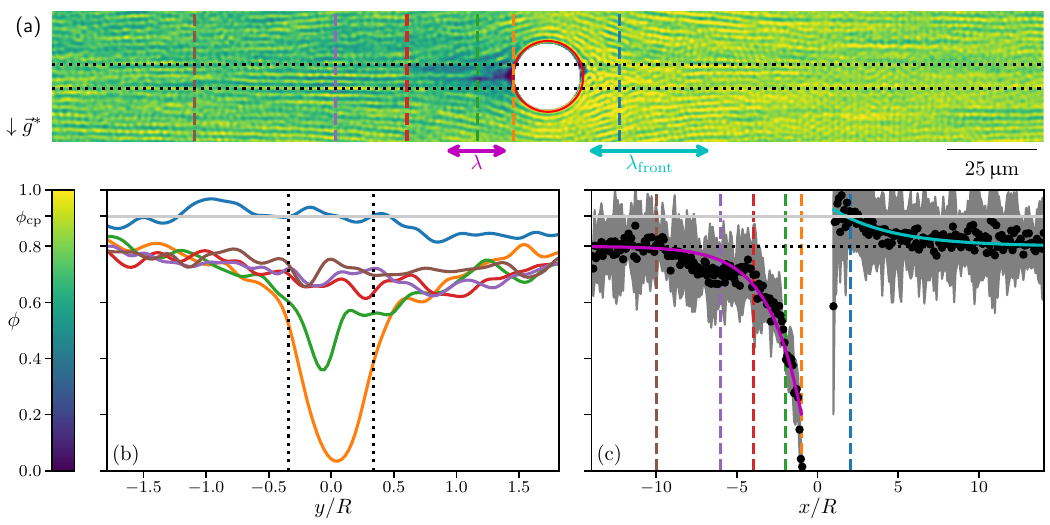}
    \caption{(a) Map of the packing fraction time-averaged in the frame of the obstacle in a passive case where $\mathrm{Pe}\approx 200$, see colour bar below. A Gaussian smoothing of size $a/2$ has been applied. The excluded zone is shown as a red circle of radius $R$. The white disk represents the obstacle of radius $R-a$. 
    (b) Profiles along $y$ at various $x$ positions behind the obstacle indicated on (a) and (c) by vertical dashed lines of corresponding colours. A Gaussian smoothing of size $a$ has been applied.
    (c) Profile along $x$ that is $y$-averaged on $|y|<2a$, i.e. dotted black lines in (a) and (b). A baseline measured at rest has been removed. The horizontal black dotted line is the density at rest at the position of the moving obstacle. The cyan and magenta curves are the best fits by exponential decay respectively in front and behind the obstacle. The respective decay lengths are shown as double arrows below (a). The horizontal gray line in (b) and (c) highlights the hard disk close packing fraction $\phi_\mathrm{cp}$.
    }
    \label{fig:xp_passive_density}
\end{figure*}

The obstacle was forged from a borosilicate rod (BR-100-15, $\varnothing\SI{1}{\milli\metre}$, \SI{15}{\centi\metre}) using a micropipette puller (P-1000 Sutter Instrument). We used two different protocols to obtain two obstacle sizes. After detachment of the two halves of the rod, we performed 30 cycles of fire polishing of the tip in order to obtain a smooth hemisphere.
The rod is held by a micromanipulator (MP-285A from Sutter Instruments), so that the obstacle can be lowered in the monolayer within \SI{0.5}{\micro\metre} from the surface of the container and then moved along an isodensity line (within \SI{3}{\degree}), as shown in  Fig.~\ref{fig:xp}(b).
Each measurement consists of two back-and-forth motions of \SI{200}{\micro\metre} at respectively $U_\mathrm{obs}=\SI{10}{\micro\meter/\second}$ and $U_\mathrm{obs}=\SI{5}{\micro\meter/\second}$, separated by a \SI{40}{\second} waiting time.
We find that a steady state is reached after \SI{100}{\micro\metre}, and thus average our data over the second half of the test. 
As the obstacle appears on the image only as a featureless dark zone contiguous to the wake, we use Euclidean distance transform to locate the obstacle and measure the apparent exclusion radius $R$ on each image. We have kept only experiments where the relative variations of $R$ within the time interval of interest are less than 5\%. However, from experiment to experiment $R$ varies between \SI{6.6}{\micro\metre} and \SI{9.2}{\micro\metre} for the small obstacle depending on the distance between the tip of the obstacle and the surface of the container.
For the large obstacle, $R$ varies between \SI{38}{\micro\metre} and \SI{50}{\micro\metre}.
We use the trackpy library~\cite{trackpy2016} to locate the position of each particle, except in the vicinity of the large obstacle where uneven light reflection prevents quantitative image analysis. 
An undisturbed density profile $\rho(y)$ is measured far from the obstacle, from which we can extract the equation of state of the suspension~\cite{piazza1993}. Fitting the divergence of this equation of state with respect to density with an analytical expression for a solid of hard disks~\cite{salsburg1962EquationStateClassical}, we extract the effective hard disk radius of the particles $a=\SI{950\pm 10}{\nano\metre}$.
We can thus define a packing fraction as $\phi=\pi a^2\rho$.
The density field is then binned on a \SI{0.273}{\micro\metre}-spaced grid in the moving frame of the obstacle and time-averaged.

Our experimental parameters are the position $y_\mathrm{obs}$ along the density gradient at which we lower the obstacle, which sets the density, and the concentration of \ce{H2O2} we add to induce self-propulsion of the particles, which sets the activity.
The self-propulsion is often quantified by comparing the persistence length $v_\mathrm{p}\tau_\mathrm{r}$ to the radius of the particle $a$, where $v_\mathrm{p}$ is the self-propulsion velocity under dilute conditions, and $\tau_\mathrm{r}$ its persistence time (inverse rotational diffusivity)~\cite{solon2015active}.
As in~\cite{theurkauff2012dynamic, ginot2015nonequilibrium}, we quantify the activity by measuring the effective sedimentation length $\ell^\mathrm{sedim}_\mathrm{eff}$ from the sedimentation profile in the dilute regime. The ratio between active and passive (no \ce{H2O2}) sedimentation lengths is then converted to 
$v_\mathrm{p} = \sqrt{\frac{2D_\mathrm{t}}{\tau_\mathrm{r}}(\ell^\mathrm{sedim}_\mathrm{eff}/\ell^\mathrm{sedim}_0-1)}$, where $\tau_\mathrm{r}=\SI{2}{\second}$ and $D_\mathrm{t}=\SI{0.36}{\micro\metre^2/\second}$ have been measured in a dilute case with vanishing $g^*$.

\section{Experimental results in passive cases}
Figure~\ref{fig:xp_size_dependency} displays typical snapshots of a moving obstacle in a passive sediment. At all considered experimental conditions, we observe a density wake with an almost empty zone behind the obstacle. However, the length of this wake changes by about two orders of magnitude in units of the particle size. For instance, in Fig.~\ref{fig:xp_size_dependency}(a), the wake is a few particle size long, less than the obstacle diameter, whereas in Fig.~\ref{fig:xp_size_dependency}(b) the wake extends over hundreds of particle size. Between the two panels, only the obstacle diameter changed significantly (by a factor of $\approx 5$) although we adjusted the scales of the pictures so that the two obstacles are displayed at the same size. This representation highlights qualitatively that the wake length responds non linearly to the size of the obstacle.

A more precise description can be obtained by time-averaging the density in the frame of the obstacle, as shown in Fig.~\ref{fig:xp_passive_density}(a). Here, we observe both the low-density wake behind the obstacle and an accumulation of particles in front. Contrary to the well-studied dilute case~\cite{squires2005simple}, in this dense regime several layers of particles stack in front of the obstacle.
More generally, the density field is structured in layers that follow the obstacle curvature in front and on the sides, but that detach behind, are advected and eventually close well-behind the obstacle.
This is qualitatively consistent with 3D experiments at high packing fractions~\cite{sriram2010ActiveMicrorheologyColloidal}, but strikingly different from recent analytical predictions in which the wake would flare out behind the obstacle before closing~\cite{banerjee2022unjamming}.

To study more quantitatively the density perturbation around the obstacle, we show in Fig.~\ref{fig:xp_passive_density}(b) transverse density profiles at various positions along the wake. 
The blue curve is one radius ahead of the obstacle where the density reaches close packing.
The orange curve is tangent to the exclusion zone just behind the obstacle and shows a deep minimum where the density goes to zero. This minimum becomes shallower (green curve) further away behind the obstacle until it disappears.

To quantify the wake more precisely, we average the density vertically over two particle diameters and plot the longitudinal density profile $\bar{\phi}(x) = (4a)^{-1}\int_{y_\mathrm{obs}-2a}^{y_\mathrm{obs}+2a}\phi(x,y)\mathrm{d}y$ shown in Fig.~\ref{fig:xp_passive_density}(c).
Immediately in front of the obstacle, we observe that the density can reach values above close packing $\bar{\phi}>\phi_\mathrm{cp}$. This is reminiscent of the jammed area observed in granular experiments~\cite{kolb2013RigidIntruderTwodimensional}.
In order to extract a length characteristic of the closing of the wake, we fit $\bar{\phi}(x)$ behind the obstacle with an exponential decay (magenta curve). We call the resulting length $\lambda$ the \emph{closing length}.
A similar procedure performed ahead of the obstacle (cyan curve) yields a second length that we call the \emph{front length}.
Determining $\lambda$ require quantitative particle tracking in the immediate vicinity of the obstacle and is thus only accessible for the small obstacle.

\begin{figure}
    \includegraphics{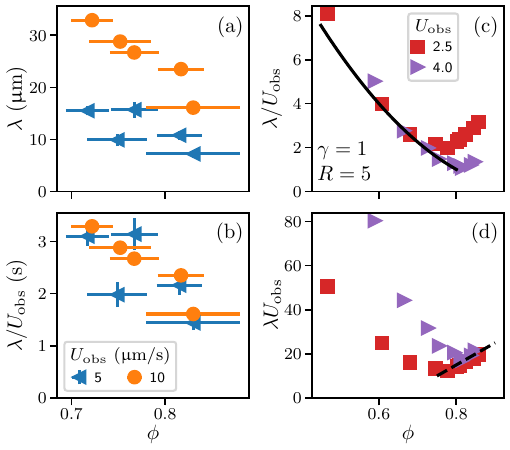}
    \caption{(a) Experimental closing length as a function of packing fraction for two obstacle velocities in the passive case for small obstacles ($\SI{8}{\micro\metre}\leq R\leq\SI{10}{\micro\metre}$).
    (b) Same data for the closing time.
    (c) Closing time from simulations as a function of packing fraction for two obstacle velocities in the passive case. Solid black curve corresponds to $\gamma R^2(1-\phi)^2$.
    (d) Same data with the inverse scaling in $U_\mathrm{obs}$. Dashed black line is an affine guide for the eye.
    }
    \label{fig:U_dependency}
\end{figure}

In Fig.~\ref{fig:U_dependency}(a), we show that the closing length in passive cases decreases with increasing packing fraction. Here the packing fraction is measured while the obstacle is at rest and averaged over the positions that the obstacle will occupy during its motion. We also observe that the closing length depends on the velocity of the obstacle. We thus construct the closing time $\tau=\lambda/U_\mathrm{obs}$ in order to collapse the data for the two obstacle velocities, as shown in Fig.~\ref{fig:U_dependency}(b).

To characterise further the scaling of $\lambda$ we will turn to numerical simulations. From this characterisation, we will sketch analytical arguments that will extract the important physical ingredients to make sense of both numerical and experimental measurements.

\section{Numerical simulations}
To extend the validity of our experimental observations to a wider range of parameters, we perform numerical simulations of active Brownian particles, which have proved to be a suitable model for Janus colloids~\cite{klongvessa2022aging,finscarreira2024HowSteerActive}. 

We thus consider $N$ active Brownian particles in a box of size $L_x \times L_y$ whose positions $\br_i$ and orientations $\bu(\theta_i) = (\cos\theta_i, \sin\theta_i)$ evolve in a two-dimensional space as
\begin{equation}\label{eq:dynamics-abp}
    \begin{split}
   \gamma \dot \br_i & = f_\mathrm{p} \bu(\theta_i) + \bF_i^{\rm WCA} \!\!+ \bF_i^{\rm obs} \! -f_g \textbf{ e}_y  \! + \sqrt{2\gamma k_\mathrm{B} T}\boldsymbol{\eta}_i \;,\\ 
    \dot \theta_i & = \sqrt{2D_r}\zeta_i  \;,
    \end{split}
\end{equation}
where $\gamma$ is the friction coefficient, $f_\mathrm{p}$ is the self-propulsion force, $D_r$ is the rotational diffusivity, $f_g$ is the magnitude of the gravitational force acting along the y direction $\textbf{ e}_y$, and $\boldsymbol{\eta}_i$ and $\zeta_i$ are centred unit-variance Gaussian white noises. In the absence of interactions the bare self-propulsion velocity is $v_\mathrm{p}=f_\mathrm{p}/\gamma$, and the passive dilute translational diffusivity is $D_t=k_\mathrm{B} T/\gamma$.
$\bF_i^{\rm WCA}$ and $\bF_i^{\rm obs}$ are respectively the inter-particle forces resulting from a Weeks-Chandler-Andersen (WCA) potential, and the repulsive force between particles and the moving obstacle, whose centre of mass evolves as
\begin{equation}\label{eq:dynamics-obstacle}
    \begin{split}
    \dot x_\mathrm{obs}(t) &= U_\mathrm{obs}\;,\\
    \dot y_\mathrm{obs}(t) &=0\;,
    \end{split}
\end{equation}
where $U_\mathrm{obs}$ is the constant obstacle velocity. As in the experiments, the obstacle moves at constant velocity rather than at constant force and the obstacle position $y_\mathrm{obs}$ is used to probe different densities. 
See the Appendix~\ref{sec:numerical_details} for definitions of the different inter-particle potentials and for simulation details. We stress that, in such simulations, the viscosity, the pressure, and the compressibility are emerging properties, whereas the friction coefficient is a parameter of the simulation.

The closing length is estimated by the same procedure as in experiments. 
In Fig.~\ref{fig:U_dependency}(c) we show the $\phi$-dependency of the closing time $\tau=\lambda/U_\mathrm{obs}$ at fixed obstacle size and friction coefficient in the passive case ($f_\mathrm{p}=0$). As in experiments, we observe a decreasing regime where the closing times at various obstacle velocities collapse.
However, at higher densities the closing time also displays an increasing regime where the different obstacle velocities do no collapse, indicating a different scaling in $U_\mathrm{obs}$.
In Fig.~\ref{fig:U_dependency}(d) we plot instead $\lambda U_\mathrm{obs}$ that displays a good collapse within the increasing regime.
We thus observe two distinct regimes in densities of possibly different physical origins. The minimum between the two regimes shifts towards higher densities when $U_\mathrm{obs}$ increases.
We also notice that the density just behind the obstacle is null ($\phi<0.01$) in the whole decreasing regime as observed in experiments. However the density starts to rise significantly above zero at the crossover, up to $\phi\approx 0.2$ far in the increasing regime (not shown).

\begin{figure}
    \includegraphics[]{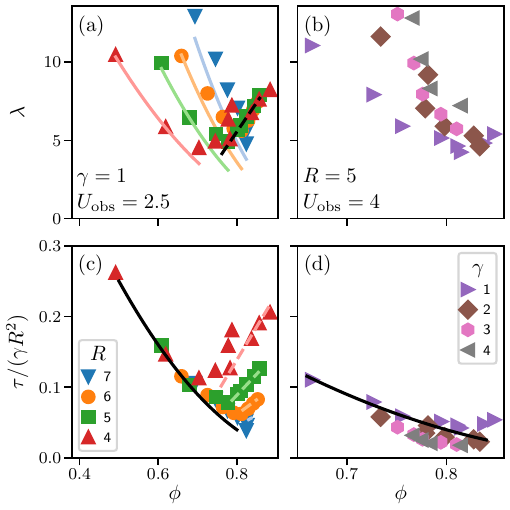}
    \caption{Closing length as a function of packing fraction in passive simulations with gravity. 
    (a) Different obstacle radii.
    (b) Different friction parameter $\gamma$.
    (c,d) Same data as (a,b) rescaled by $U_\mathrm{obs}\gamma R^2$.
    Solid black line is $(1-\phi)^2$ that is the inverse compressibility factor  of a hard disk fluid. It corresponds to the solid coloured lines in (a). 
    Dashed black line in (a) is an affine guide for the eye, it corresponds to the dashed coloured lines in (c).
    }       
    \label{fig:abp_R_friction_dependency}
\end{figure}

In Fig.~\ref{fig:abp_R_friction_dependency}(a) we show the $\phi$-dependency of $\lambda$ at fixed obstacle velocity and friction coefficient in the passive case ($f_\mathrm{p}=0$). 
For all radii, we observe the decreasing regime followed by an increasing one.
The former shifts toward higher densities as $R$ increases, whereas the latter is independent of $R$.
When rescaled by $R^2$, as shown in Fig.~\ref{fig:abp_R_friction_dependency}(c), the various values of $R$ collapse in the decreasing regime. 
In Fig.~\ref{fig:abp_R_friction_dependency}(c), we confirm that increasing the obstacle radius pushes the transition between the two regimes towards higher packing fractions. 
In Fig.~\ref{fig:abp_R_friction_dependency}(b), we show that the increasing regime disappears from our accessible packing fraction window as we increase the friction. In Fig.~\ref{fig:abp_R_friction_dependency}(d), we show that the curves collapse on the decreasing regime once scaled by $\gamma$.

In summary, in the decreasing regime the closing length scales as $\lambda\propto \gamma U_\mathrm{obs} R^2$ times a decreasing function of $\phi$, whereas in the increasing regime the closing length scales as $1/U_\mathrm{obs}$ and is independent of $R$.
The decreasing regime is characterised by a cavity (quasi zero density) behind the obstacle, whereas in the increasing regime the density is significantly above zero behind the obstacle.
In the following, we will use these observations to extract the physics underlying both regimes.

\section{Theoretical arguments}

\paragraph{Continuum description.}
We now rationalise the observed regimes for the closing length.
Let us describe the passive colloidal bath by its density $\rho(\br,t)$, its pressure $P(\rho)$, and the friction coefficient $\gamma$ of the colloids with the substrate and the surrounding fluid.
Particle number is conserved, $\partial_t\rho=-\nabla\cdot\bj$, and in the overdamped regime the particle current is driven by pressure gradients against friction,
\begin{equation}\label{eq:current}
    \bj=\rho\bv=-\gamma^{-1}\nabla P .
\end{equation}
A pressure difference $\Delta P$ acting over a distance $\ell$ thus moves the colloids at a velocity $v\sim\Delta P/(\gamma\rho\,\ell)$.
This single estimate is all we need below; the only question is which pressure difference and which distance are relevant.

\paragraph*{Two lengths and a Péclet number.}
Two lengths enter the problem.
The first is structural: it is the obstacle radius $R$.
The second is dynamical: it is the distance $\ell_U$ over which the largest possible depression, \emph{i.e.} $\Delta P = P-0$, creates a flow that is just fast enough to keep up with the obstacle, $P/(\gamma\rho\,\ell_U)\sim U_\mathrm{obs}$, that is
\begin{equation}\label{eq:lU}
    \ell_U=\frac{P}{\gamma\rho U_\mathrm{obs}}=\frac{k_\mathrm{B}T Z}{\gamma U_\mathrm{obs}}\;,
\end{equation}
where the compressibility factor $Z = P/(\rho k_\mathrm{B}T)$, which quantifies the deviation from ideal gas behaviour.
Equivalently, $\ell_U=D/U_\mathrm{obs}$ is the usual advection length built on an effective collective diffusion coefficient $D=P/(\gamma\rho)$.
The ratio of the two lengths defines a Péclet number,
\begin{equation}\label{eq:Pe}
    \mathrm{Pe}=\frac{R}{\ell_U}=\frac{\gamma\rho U_\mathrm{obs}R}{P}=\frac{\gamma U_\mathrm{obs}R}{k_\mathrm{B}T\,Z}\;,
\end{equation}
which can equivalently be read as the ratio of the obstacle velocity to the refilling velocity $P/(\gamma\rho R)$ that the full bath pressure imposes over the obstacle size.
For systems of passive particles interacting via repulsive interactions, $Z$ increases with density, thus $\mathrm{Pe}$ decreases with density.
In the dilute limit, $Z=1$ and $k_\mathrm{B}T/\gamma=D_\mathrm{t}$, so that Eq.~(\ref{eq:Pe}) reduces to the Péclet number $\mathrm{Pe}^\mathrm{dilute}$ introduced above; in dense suspensions it is smaller by the factor $Z$.

\paragraph*{Large Péclet number: refilling a cavity.}
For $\mathrm{Pe}\gg1$ the obstacle moves faster than the bath can follow, and it leaves behind it an empty region of width $\sim R$, as observed in Fig.~\ref{fig:xp_passive_density}.
The bath then pushes colloids into this void with the full pressure $\Delta P=P$ acting over the distance $R$, \textit{i.e.} at velocity $v\sim P/(\gamma\rho R)$, and the cavity closes after a time $\tau\sim R/v$.
The corresponding closing length $\lambda=U_\mathrm{obs}\tau$ is
\begin{equation}\label{eq:void}
    \lambda\sim R\,\mathrm{Pe}=\frac{\gamma\rho U_\mathrm{obs} R^2}{P}
    = \frac{\gamma U_\mathrm{obs} R^2}{k_\mathrm{B}T Z}.
\end{equation}
It reproduces the signatures of the decreasing regime: $\lambda$ is proportional to $\gamma U_\mathrm{obs}R^2$ and decreases with density through $1/Z$.
The solid black curves in Fig.~\ref{fig:U_dependency}(c) and Fig.~\ref{fig:abp_R_friction_dependency}(c,d) represent Eq.~(\ref{eq:void}) using the compressibility factor of a hard disk fluid $Z=1/(1-\phi)^2$~\cite{helfand1961TheoryTwoOne}, which is in good agreement with the data for packing fractions below 0.69, where a fluid state is expected.

\paragraph*{Small Péclet number: no cavity.}
For $\mathrm{Pe}\ll 1$ the bath refills faster than the obstacle opens a gap. 
The density perturbation does not reach zero behind the obstacle.
Nevertheless, we will assume that the pressure just behind the obstacle is negligible with respect to the bath pressure $P$, thus $\Delta P\approx P$. This assumption is justified at high packing where $P$ is a steep non linear function of density.
In this regime the extent of the wake is $\ell_U$ itself

\begin{equation}\label{eq:lambda_increasing}
    \lambda\sim \frac{R}{\mathrm{Pe}}=\frac{P}{\gamma\rho U_\mathrm{obs}}
    =\frac{k_\mathrm{B}T\,Z}{\gamma U_\mathrm{obs}}.
\end{equation}
The scaling of Eq.~\eqref{eq:lambda_increasing} is compatible with the observations of the numerical simulations in the increasing regime: 
$\lambda$ is independent of $R$, scales as $1/U_\mathrm{obs}$ and increases with density through $Z$.

\paragraph{Crossover.}
Equations~(\ref{eq:void}) and (\ref{eq:lambda_increasing}) can be summarised as $\lambda/R\sim\mathrm{Pe}$ for $\mathrm{Pe}\gg1$ and $\lambda/R\sim1/\mathrm{Pe}$ for $\mathrm{Pe}\ll1$: the closing length is always larger than the obstacle, and is minimal, $\lambda\approx R$, at $\mathrm{Pe}\approx1$, corresponding to a compressibility factor $Z_\mathrm{c} = \gamma U_\mathrm{obs}R/(k_\mathrm{B}T)$.
As the density increases, $\mathrm{Pe}$ decreases and the system goes from the first regime to the second, which explains the minimum of $\lambda$ in Fig.~\ref{fig:U_dependency} and Fig.~\ref{fig:abp_R_friction_dependency}.

\paragraph{Relation to previous works.}
When the obstacle only weakly perturbs the bath, Eq.~(\ref{eq:current}) can be linearised about the mean density $\bar\rho$, giving a diffusion equation for the density perturbation with the collective diffusion coefficient $P^\prime(\bar\rho)/\gamma$, with $P^\prime=\mathrm{d}P/\mathrm{d}\rho$~\cite{demery2014generalized, poncet2017universal, banerjee2022unjamming}.
The same two lengths then appear, with $P/\rho$ replaced by $P^\prime(\bar\rho)$: a ``diffusive wake'' behind the obstacle, of length $U_\mathrm{obs}\gamma R^2/P^\prime(\bar\rho)$, and a perturbation in front of it, of extent $P^\prime(\bar\rho)/(\gamma U_\mathrm{obs})$~\cite{banerjee2022unjamming}.
Our argument differs in two respects.
First, we consider regimes where the pressure behind the obstacle is either zero (large $\mathrm{Pe}$) or negligible with respect to the bath pressure (low $\mathrm{Pe}$ and high packing), so that the linearisation is not justified. Refilling is then driven by the full pressure $P$ rather than by the stiffness $P^\prime$, which turns $P^\prime(\bar\rho)$ into $P/\rho$ in Eq.~(\ref{eq:void}).
In other words, the relevant collective diffusion coefficient is $Z D_\mathrm{t}$ instead of $(\mathrm{d}\phi Z)/(\mathrm{d}\phi) D_\mathrm{t}$ as in~\cite{sriram2010ActiveMicrorheologyColloidal}.
Second, the length of Eq.~(\ref{eq:lambda_increasing}) was obtained by \citet{banerjee2022unjamming} for the perturbation \emph{in front} of the obstacle, while the perturbation behind is always described by the diffusive scaling of Eq.~\eqref{eq:void}.
By contrast, we find that Eq.~(\ref{eq:lambda_increasing}) also controls the wake \emph{behind} the obstacle at small $\mathrm{Pe}$, where the diffusive wake, of length $R\,\mathrm{Pe}<R$, is hidden by the obstacle itself.
This length cannot be recovered from the dipolar density perturbation found by \citet{squires2005simple} at small Péclet number.
Furthermore, we note that in Fig.~\ref{fig:U_dependency}(d) and Fig.~\ref{fig:abp_R_friction_dependency}(a), the dashed black lines hint toward $\lambda\propto \phi$ in the increasing regime, which would imply $Z\propto\phi$ in Eq.~(\ref{eq:lambda_increasing}). This is consistent with phenomenological observations of \citet{banerjee2022unjamming} but not with the equilibrium equation of state of the fluid or of the crystal of hard disks.

\section{Compressibility factor}
Our study on passive systems, in particular Eq.~(\ref{eq:void}) and Eq.~(\ref{eq:lambda_increasing}), suggests that the compressibility factor is the natural variable to describe the density dependence. However, our data suggest that we cannot rely on analytical expressions for hard disks in the dense regime.
Furthermore, in active systems, the function $Z(\phi)$ depends on activity and can display non-monotonic behaviour as a function of $\phi$ for high enough activities~\cite{ginot2015nonequilibrium, cates2015motility, solon2015PressureNotState}. Thus, in order to study the effect of activity, we ought to plot our data as a function of $Z$ rather than as a function of $\phi$. To this end, in both our simulations under gravity and in our experiments, we measure the compressibility factor $Z$ directly at the altitude $y_\mathrm{obs}$ of the obstacle from the sedimentation profile. Following~\cite{piazza1993}, we define the hydrostatic pressure from an integral of the density $\rho(y)$ estimated at rest or far from the obstacle:
\begin{equation}
    P(y_\mathrm{obs}) = mg^*\int_{y_\mathrm{obs}}^{\infty} \rho(y)\mathrm{d}y,
\end{equation}
where $m$ is the buoyant mass of a particle, $g^*$ the in-plane gravity. In terms of packing fraction, the compressibility is then
\begin{equation}
    Z(y_\mathrm{obs}) = \frac{1}{\ell_\mathrm{g}\phi(y_\mathrm{obs})}\int_{y_\mathrm{obs}}^{\infty} \phi(y)\mathrm{d}y.
    \label{eq:Z_from_profile}
\end{equation}
We measure the sedimentation length $\ell_\mathrm{g} = k_\mathrm{B}T/(mg^*)$ at the top of the passive sedimentation profile where the dilute sediment behaves as an ideal gas and $\phi\propto\exp(-y/\ell_\mathrm{g})$.
In practice, we integrate only up to a threshold altitude $y_\mathrm{max}$ where the density is low enough to be in the dilute regime but high enough to be estimated reliably. The integral in $[y_\mathrm{max},\infty]$ is estimated from the dilute exponential law and is negligible at the densities considered in the present study.
Previous studies have shown that the compressibility factor of active Janus colloids or ABPs can be obtained from the same method~\cite{ginot2015nonequilibrium}.

\begin{figure}
    \includegraphics{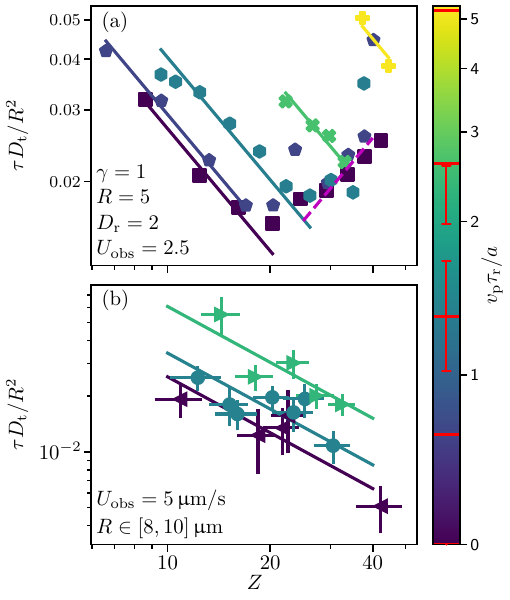}
    \caption{Effect of activity. Rescaled closing time as a function of the measured compressibility factor at different activities for (a) simulations and (b) experiments. The rescaling is according to the decreasing regime (solid lines), since $\gamma$, $R$ and $U_\mathrm{obs}$ are constant in (a) a collapse of the increasing regime is also expected (dashed magenta line). On the colour bar, persistence lengths of the simulations are shown as red horizontal lines, whereas persistence lengths of the experiments are shown as vertical error bars.}
    \label{fig:activity_dependence}
\end{figure}

\section{Active systems}
In Fig~\ref{fig:activity_dependence}(a) we show the $Z$-dependence of the closing time at various activities but at the same friction, obstacle radius and velocity. In the passive case, we confirm the $1/Z$ dependence in the decreasing regime (dark blue solid line) and the $Z$ dependence in the increasing regime (dashed magenta line). With increasing activity, we observe that the increasing regime remains unchanged, but that the decreasing regime shifts up, and the transition between the two regimes shifts towards larger values of $Z$. 
Experimentally, we observe only the decreasing regime, as shown in Fig~\ref{fig:activity_dependence}(b). We attribute the absence of increasing regime to a higher translational friction in our experiments due to the proximity of the colloidal particles to the bottom of the container. As seen in Fig.~\ref{fig:abp_R_friction_dependency}(b,d), larger friction can push the increasing regime to inaccessibly high densities. Nevertheless, we confirm experimentally that the decreasing regime shifts up with activity.

\begin{figure}
    \includegraphics{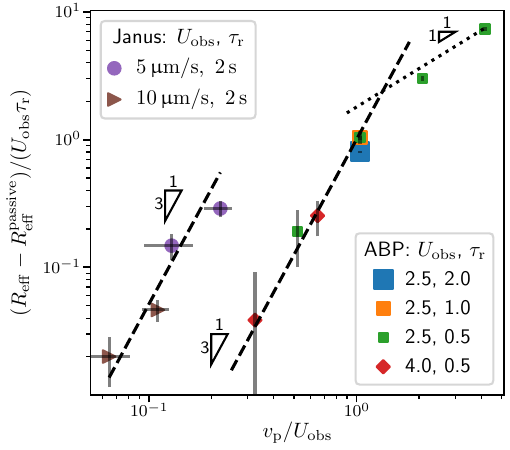}
    \caption{Increase of effective obstacle radius with activity in both experiments (Janus) and simulations (ABP, $R=5$), scaled by the distance moved by the obstacle during the persistence time of the particles, as a function of the ratio between particle velocity and obstacle velocity. Dashed lines are power laws of exponent 3. Dotted line is a power law of exponent 1.}
    \label{fig:Reff_activity}
\end{figure}

Referring to Eq.~(\ref{eq:void}), a shift up in the decreasing regime can be due either to a decrease in effective temperature, or to an effective increase of the friction, the obstacle velocity or the obstacle radius. However, according to Eq.~(\ref{eq:lambda_increasing}) any of these changes would also affect the increasing regime, except for the radius $R$. Therefore the effects of activity are only compatible with an effectively larger obstacle. We thus fit the decreasing regime at each activity to extract the effective obstacle radius $R_\mathrm{eff}$. 
Extending arguments of \citet{burkholder2020NonlinearMicrorheologyActivea} outside of the dilute regime, we expect that active effects become important when $v_\mathrm{p}$ becomes comparable to or larger than $U_\mathrm{obs}$. We thus plot the increment of $R_\mathrm{eff}$ with respect to the passive case as a function of $v_\mathrm{p}/U_\mathrm{obs}$ in Fig.~\ref{fig:Reff_activity}. In order to collapse simulations performed at different persistence times, we have to scale the effective radius increment with $U_\mathrm{obs}\tau_\mathrm{r}$, i.e. the distance moved by the obstacle during a persistence time of the particles.
Empirically, we found that numerical data display a crossover between a linear regime for $v_\mathrm{p}\gg U_\mathrm{obs}$ and a cubic regime for $v_\mathrm{p}\ll U_\mathrm{obs}$. Although the persistence length range explored in experiments is comparable to simulations, experimental data explore only the $v_\mathrm{p}\ll U_\mathrm{obs}$ and display the cubic regime with a different prefactor compared to simulations.

Quantitative disagreement could be due to an underestimation of either $v_\mathrm{p}$ or $\tau_\mathrm{r}$ by a factor $\approx 2$, or to physical ingredients absent from the ABP simulations: 3D nature of the rotational diffusion~\cite{hagen2011BrownianMotionSelfpropelled}, hydrodynamic or phoretic alignment of particle polarity with the surface of the probe~\cite{finscarreira2024HowSteerActive}, or hydrodynamic interactions within the suspension~\cite{yanagishima2021ParticleLevelVisualizationHydrodynamica}. 

\section{Conclusion and perspectives}
We have quantified the length of the density wake at short range behind a moving obstacle in a dense monolayer of active colloidal particles in experiments and numerical simulations. Analytical arguments from a passive hydrodynamic model taking into account friction and compressibility of the suspension enabled us to identify two regimes where the length of the wake successively decreases and then increases with density. Experiments with passive particles are consistent with the decreasing regime where the physics is governed by the presence of an empty region just behind the obstacle, in particular the non linear scaling in obstacle size.

We found that endowing the particles with persistent self-propulsion has little effect on the increasing regime (higher densities) but that it increases the length of the wake in the decreasing regime (lower densities). The effect of activity on the length of the wake in both decreasing and increasing regimes can be understood in terms of an effectively larger obstacle. When the obstacle is slow with respect to the self-propulsion of the particles, we found an increment in effective radius proportional to the distance moved by the obstacle during the persistence time of the particles. In the decreasing regime, the effect of activity leads to a longer wake than in the passive case at constant compressibility.
This finding goes against previous analytical and numerical investigations in the dilute regime that found a faster ballistic closing of the wake for active systems at constant density~\cite{burkholder2020NonlinearMicrorheologyActivea, knezevic2021OscillatoryActiveMicrorheologya}. We attribute this apparent discrepancy to the fact that the compressibility factor $Z$ controls wake closing and that $Z$ depends on both density and activity. 
At low density where $Z$ increases with the propulsion velocity~\cite{cates2015motility}, we expect a decrease in wake length that could mask a modest increment in effective radius directly due to activity.

To sum up, the density wake behind an obstacle is a sensitive measurement of the compressibility of a suspension of passive or active particles, and its dependence on density. In particular at densities where the suspension at rest would be crystalline, the movement of the obstacle continuously destroys the crystalline order, allowing a measurement of the compressiblity of the amorphous system behind the obstacle.
However, the rear of the obstacle is not the whole story.
In front of the obstacle we observe large densities, even higher than close packing. Here, we predict a much higher viscosity, or even an elasto-plastic behaviour that might dominate the shape of the density perturbation in front.
More generally, our study has characterised a part of the density perturbation around a moving obstacle, that is not as simple as in the dilute case. From the whole density field, one could infer the average force acting on the obstacle and progress toward an understanding of the microrheology of passive and active suspensions.

\begin{acknowledgments}
We thank Shashi Thutupalli, Sriram Ramaswamy, and Patrice Bacchin for useful discussions. ML and CCB acknowledge support from the ANR ActivSoliPolar (ANR-24-CE30-6841). GD, CCB and HDA acknowledge support from the MITI Rheodaco. GS acknowledges support from the ERC Advanced Grant ActBio (funded as UKRI Frontier Research Grant EP/Y033981/1).
VD dedicates this work to the memory of Rudi Podgornik, a mentor and friend.
\end{acknowledgments}

\section*{Author declarations}
\paragraph{Conflict of interest} None.

\paragraph{Author contributions}
Investigation (simulations) by GS (lead).
Investigation (experimental) by GD (lead) and BB.
Data curation by GD, GS, BB, and ML.
Formal analysis and software by GD, GS, HDA, and ML.
Methodology by VD, HDA, and ML.
Conceptualization, funding acquisition, project administration, supervision by FG, HDA, CCB and ML.
Validation by HDA, CCB, and ML.
Visualisation: ML.
Writing original draft : GS, FG, VD, CCB, HDA and ML (lead).
Writing review \& editing :
 
\paragraph{Data availability statement}
Data for this article, including density maps from experiments and simulations, tables of intermediate results as well as the \texttt{Matplotlib}~\cite{Hunter:2007} scripts to generate the figures are available at Zenodo at \url{https://doi.org/10.5281/zenodo.23480622}

\appendix* 
\section{Numerical details}\label{sec:numerical_details}

To integrate the dynamics corresponding to Eqs.~(5-6), we use a Euler scheme with adaptive time steps. At each iteration, the time step is set to $\Delta t = \min \big [ \Delta t_{\rm max}, \frac{r_0}{10 F_{\rm M}} \big ]$, with $r_0=1$ being the interaction range of the WCA interparticle potential, $ F_{\rm M}=\max_i\left\{|F^{\rm int}_{i,x}|,|F^{\rm int}_{i,y}|\right\}$ is the largest absolute Cartesian component of the interparticle force acting on any particle at the current time step, and $\Delta t_{\rm max}=10^{-2}$ for passive simulations and $\Delta t_{\rm max}=10^{-3}$ for active simulations with $v_0\ge 5$.\\  
Interparticle forces ${\bf F}_i^{\rm WCA}= - \sum_{j} \nabla V^{\rm WCA}(\br_j -\br_i) $ results from a  Weeks-Chandler-Andersen (WCA)\begin{equation}\label{eq:potential}
V^{\rm WCA}(r) =
\begin{cases}
  4 \varepsilon \left[ \left( \frac{\sigma}{r} \right)^{12} - \left( \frac{\sigma}{r} \right)^{6} \right] & \mathrm{if} \quad r < r_0 \;, \\
  0        & \mathrm{if} \quad r \geq r_0 \; ,
\end{cases}
\end{equation}
where $\varepsilon$ is the depth of the potential, $\sigma$ is the particle diameter, and $r_0 = 1$ is the interaction range. We use $\sigma= r_0/2^{1/6} $ to retain only the repulsive part. 
The obstacle is modeled as a diverging radial potential
\begin{equation}\label{eq:potential-obs}
  V^\mathrm{obs}(r=|\mathbf{R} - \mathbf{r}_i|) =
  \begin{cases}
    \frac{\Omega}{\nu} (R_\mathrm{obs} - r)^{\nu }  & \mathrm{if} \quad r < R_\mathrm{obs} \;, \\
    0        & \mathrm{if} \quad r \geq R_\mathrm{obs} \; ,
  \end{cases}
\end{equation}
with $\Omega=10$ and $\nu=3$. 
Furthermore, particles can experience a gravitational force $f_g$ to further match the experiments. Since particles experience a gravitational force $f_g$, we add a confining potential along the $y$ axis to avoid vertical currents. Particle are thus confined with a confining potential 
\begin{equation} \label{eq:potential-wall}
  V^\mathrm{wall}(y) =
  \begin{cases}
    \frac{k}{p} (y_\mathrm{bottom} - y) ^{p }  & \mathrm{if} \quad y < y_\mathrm{bottom} \;, \\
    \frac{k}{p} (y- y_\mathrm{top}) ^{p }  & \mathrm{if} \quad y > y_\mathrm{top} \;, \\
    0        & \mathrm{otherwise} \; ,
  \end{cases}
\end{equation}
with $k=1$,  $p=5$, $y_{\rm bottom} = 10$, and $y_{\rm top}=Ly-10$.
We initialize $N=\rho_0 L_xL_y$ particles in a box of sizes $L_x=150$ and $L_y = 200$ with an initial density $\rho_0=1.0$, and we let the particle relax before setting the obstacle into motion. Other simulation parameters are: $\varepsilon=1$, $k_{\rm B} T=0.2$, and $f_g=0.1$.

\bibliography{biblio}

\end{document}